\documentclass[floatfix,aps,prl,showpacs,amsmath,nofootinbib,preprintnumbers,twocolumn]{revtex4}

\usepackage{graphicx}
\usepackage{bm}

\newcommand{\figurewidthw}{4.8in}
\newcommand{\figurewidth}{2.18in}
\newcommand{\figurewidths}{2.1in}

\def\({\left(}
\def\){\right)}
\def\[{\left[}
\def\]{\right]}

\def\e{\begin{equation}}
\def\q{\end{equation}}
\def\m{\begin{eqnarray}}
\def\n{\end{eqnarray}}

\begin{document}

\title{Probing the primordial Universe from the low-multipole CMB data}

\author{Cheng Cheng and Qing-Guo Huang}\email{huangqg@itp.ac.cn}
\affiliation{State Key Laboratory of Theoretical Physics, Institute of Theoretical Physics, Chinese Academy of Science, Beijing 100190, People's Republic of China}

\date{\today}

\begin{abstract}

Since the temperature fluctuations in cosmic microwave background (CMB) on large-angular scales probe length scales that were super-horizon sized at photon decoupling and hence insensitive to microphysical processes, the low-multipole CMB data are supposed to be a good probe to the physics of the primordial Universe. In this letter we will constrain the cosmological parameters in the base $\Lambda$CDM model with tensor perturbations by only using low-multipole CMB data, including Background Imaging of Cosmic Extragalactic Polarization (B2), Planck released in 2013 (P13) and Wilkinson Microwaves Anisotropy Probe 9-year data (W9). We find that a red tilted power spectrum of relic gravitational waves is compatible with the data, but a blue tilted power spectrum of scalar perturbations on the large scales is preferred at around $2\sigma$ confidence level.

\end{abstract}


\maketitle


The cosmic microwave background (CMB) is the oldest light in the universe, dating to the epoch of photon decoupling. Since CMB encodes important physics about the Universe, the precise measurements of CMB are critical to cosmology.  See a nice review about CMB in \cite{Dodelson:2003ft}.
In particular, the temperature fluctuations measured between two points separated by a large angle $(\gtrsim 1^\circ)$ arise mainly due to the difference in the gravitational potential between the two points on the last-scattering surface. This is known as the Sachs-Wolfe effect \cite{Sachs:1967er}. On such large scales any causal effects have not had time to operate. Considering that the evolution of gravitational potential caused by the dark energy which becomes dominant in the late-time Universe, the integrated Sachs-Wolfe effect \cite{Sachs:1967er} should not be completely ignored. Once we go to the angular scales below the Sachs-Wolfe plateau, the $C_{\ell}$ curves depend sensitively on a lot of microphysics characterized by a large number of parameters, such as the total mass of active neutrinos ($\sum m_\nu$), the number of relativistic species ($N_{\rm eff}$), the gravitational lensing, the abundance of light elements and so on.

Nowadays inflation \cite{Guth:1980zm,Linde:1981mu,Albrecht:1982wi} becomes the dominant paradigm for the early Universe. 
Not only does it solve the puzzles in the hot big band model, such as flatness problem, horizon problem and so on, it provides a causal origin of the density perturbations even on the large scales that were outside the horizon at the time of recombination. 
According to the general relativity, there are three kinds of perturbations, namely scalar, vector and tensor (gravitational waves) perturbations. At the linear order, these three kinds of perturbations evolve independently and therefore we can analyze them separately. Since there are no rotational velocity fields during inflation, the vector perturbations are not excited. Thus we only need to consider the scalar and gravitational waves perturbations.

An adiabatic, Gaussian and nearly scale-invariant power spectrum of scalar perturbations has been confirmed by many cosmological observations, such as Wilkinson Microwaves Anisotropy Probe 9-year data (W9) \cite{Hinshaw:2012aka} and Planck released in 2013 (P13) \cite{Ade:2013zuv}. Actually the gravitational waves can make contributions to the temperature and polarization power spectra in
the CMB \cite{Grishchuk:1974ny,Starobinsky:1979ty,Rubakov:1982,Crittenden:1993ni,Kamionkowski:1996zd,Kamionkowski:1996ks,Hu:1997mn,Challinor:2004bd,Challinor:2005ye} as well. In the last decades, many groups tried their best to hunt for the signal of gravitational waves. Even though some hints on it were revealed in CMB in \cite{Zhao:2010ic,Zhao:2014rna}, the statistic significances were quite low (around $1\sigma$ confidence level). 
Since the relic gravitational waves damp significantly inside the horizon, ones could only expect to find the relic gravitational waves on the very large scales. Recently a breakthrough is the discovery of the relic gravitational waves by Background Imaging of Cosmic Extragalactic Polarization (B2) \cite{Ade:2014xna} where an excess of B-mode power over the base lensed-$\Lambda$CDM expectation in the range of $30\lesssim \ell \lesssim150$ multipoles. B2 found that 
\m
r=0.20_{-0.05}^{+0.07}, 
\n
and $r=0$ is disfavored at $7.0\sigma$, where $r$ is the so-called tensor-to-scalar ratio which is nothing but the ratio between the amplitudes of relic gravitational waves power spectrum and scalar power spectrum. However in 2013 Planck collaboration claimed that there is no signal for the relic gravitational waves at all, and the upper bound on the tensor-to-scalar ratio is given by $r<0.11$ at $95\%$ confidence level \cite{Ade:2013zuv} in the base six-parameter $\Lambda$CDM model where a power-law scalar power spectrum is assumed. 
Obviously there is a strong tension between B2 \cite{Ade:2014xna} and P13 \cite{Ade:2013zuv}. 
Considering the CMB spectra generated by relic gravitational waves are significant only on the large scales, in \cite{Zhao:2014rna} we combined the low-$\ell$ TT spectrum from P13 and TE spectrum from W9, and found that $r>0$ is preferred at more than $68\%$ confidence level and the maximized likelihood value of $r$ is around 0.2 which is compatible with B2. A possible explanation is that the apparent tension between B2 and P13 is coming from the the ``wrong" theoretical model, the base $\Lambda$CDM model with a power-law scalar power spectrum, adopted by Planck collaboration.

In order to reduce the possible biases from other complicated microphysics, we propose to only utilize the low-$\ell$ CMB data to probe the physics of primordial Universe because the large-angle anisotropies in CMB are not affected by any microphysics at the time of recombination. In this letter we will consider two combinations of CMB data with $\ell_{\rm max}=150$: \\
$\bullet$ one is B2+W9 (including EE, EB and BB from B2 and TT and TE from W9); \\
$\bullet$ the other is B2+P13+WP (including EE, EB and BB from B2, TT from P13 and TE from W9). \\
Here the $\Lambda$CDM model with tensor perturbations is adopted. 
The physics of primordial Universe are assumed to be encoded in both power spectra of the scalar and relic gravitational waves perturbations which are respectively parameterized by  
\m
P_s(k)&=& A_s \({k\over k_p}\)^{n_s-1}, \label{ps} \\
P_t(k)&=& rA_s \({k\over k_p}\)^{n_t}, \label{pt}
\n
where $n_s$ and $n_t$ are the spectral indices of the scalar and relic gravitational waves spectra respectively. In this letter the pivot scale is fixed to be $k_p=0.004$ Mpc$^{-1}$.
Since $\ell_{\rm max}=150$ and the data do not cover a wide perturbation modes, the power-law spectra of both scalar and relic gravitational waves in Eqs.~(\ref{ps}) and (\ref{pt}) are assumed to be applicable. 
The other free cosmological parameters are the baryon density today $(\Omega_b h^2)$, the cold dark matter density today $(\Omega_c h^2)$, the $100\times$ angular scale of the sound horizon at last-scattering ($100\theta_{\rm MC}$) and the Thomson scattering optical depth due to the reionization $(\tau)$. 

\begin{figure*}[hts]
\begin{center}
\includegraphics[width=\figurewidthw]{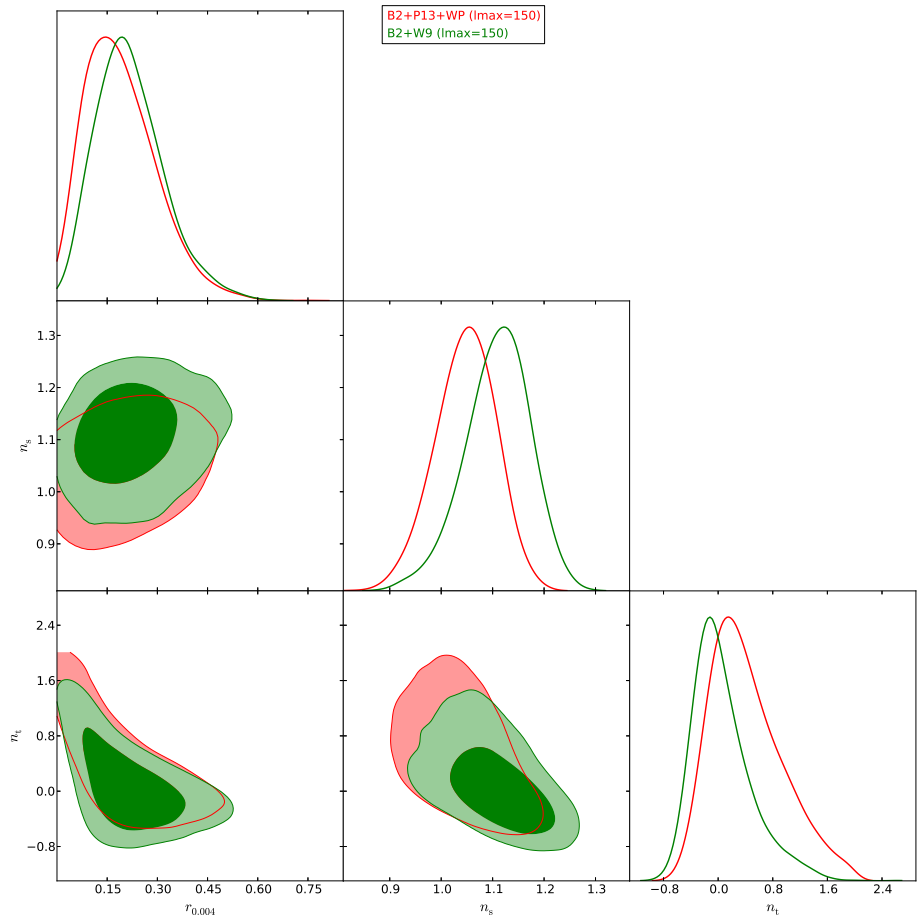}
\end{center}
\caption{The contour plots for $r$, $n_s$ and $n_t$ from low-$\ell$ CMB data in the $\Lambda$CDM+r model with $n_t$ free.
}
\label{fig:ntfree}
\end{figure*}

First of all, we take the tilt of relic gravitational wave spectrum as a free parameter $n_t$. We run the CosmoMC \cite{cosmomc} to fit the eight free running cosmological parameters, namely $\{\Omega_bh^2, \Omega_ch^2, \theta, \tau, A_s, n_s, r, n_t\}$. Our results show up in Table \ref{tab:ntfree} and Fig.~\ref{fig:ntfree}. 
\begin{table}[htbp]
\centering
\renewcommand{\arraystretch}{1.5}
\scriptsize 
{
 
\

\begin{tabular}{c|cc|cc}
\hline\hline
$n_t$ free & \multicolumn{2}{|c|}{B2+W9 $(\ell_{\rm max}=150)$} & \multicolumn{2}{c}{B2+P13+WP $(\ell_{\rm max}=150)$} \\
\hline
parameters& best fit  &$68\%$ limits  & best fit &$68\%$ limits  \\
\hline
$\Omega_b h^2$ & 0.0235 & $0.0248_{-0.0113}^{+0.0070}$ & 0.0283 & $0.0264_{-0.0141}^{+0.0078}$\\
$\Omega_c h^2$ & 0.159 & $0.160_{-0.045}^{+0.033}$ & 0.157 & $0.141_{-0.030}^{+0.018}$ \\ 
$100\theta_{\rm MC}$ & 1.145  &  $1.108_{-0.038}^{+0.050}$ & 1.140 &  $1.100_{-0.026}^{+0.044}$\\
$\tau$ & 0.102 & $0.097_{-0.018}^{+0.015}$ & 0.109 & $0.098_{-0.018}^{+0.015}$\\
$\ln(10^{10}A_s)$ & 3.058 & $3.068_{-0.059}^{+0.071}$ & 3.035 &  $3.058_{-0.053}^{+0.068} $\\
$n_s$ & 1.117 & $1.109_{-0.056}^{+0.070}$ & 1.082 & $1.047_{-0.054}^{+0.065}$\\
$r$& 0.22  & $0.22_{-0.12}^{+0.08}$  & 0.16 & $0.20_{-0.14}^{+0.07}$\\
$n_t$& 0.01  & $0.07_{-0.51}^{+0.26}$  & 0.44 & $0.43_{-0.67}^{+0.36}$ \\
\hline
\end{tabular}
}
\caption{Constraints on the cosmological parameters from low-$\ell$ CMB data in the $\Lambda$CDM+r model with $n_t$ free.  }
\label{tab:ntfree}
\end{table}
From Table \ref{tab:ntfree} and Fig.~\ref{fig:ntfree}, we see that the constraints on the cosmological parameters from the low-$\ell$ B2+W9 is consistent with those from low-$\ell$ B2+P13+WP. 
A red tilted power spectrum of relic gravitational waves is compatible with the combinations of both B2+W9 and B2+P13+WP with $\ell_{\rm max}=150$. It is consistent with our previous results in \cite{Cheng:2014bma} where only B2 data are adopted. In \cite{Cheng:2014ota} we argued that B2 provides the best choice to constrain the tilt of relic gravitational waves spectrum. Here $\ell_{\rm max}=150$ corresponds to the angular scale around $1^\circ$ in the CMB surface and the CMB spectra around $\ell_{\rm max}=150$ have been a little bit biased by the microphysics at the recombination. Compared to \cite{Cheng:2014bma}, the statistic significance for relic gravitational waves characterized by the tensor-to-scalar ratio is reduced, and a large positive $n_t$ seems also allowed when $r$ goes to around zero in this letter. Physically it implies that there are no relic gravitational waves. This region should not response to the real physics because it conflicts with the fact that there is strong evidence for the relic gravitational waves from non-zero BB spectrum detected by B2. 

\begin{figure*}[hts]
\includegraphics[width=\figurewidth]{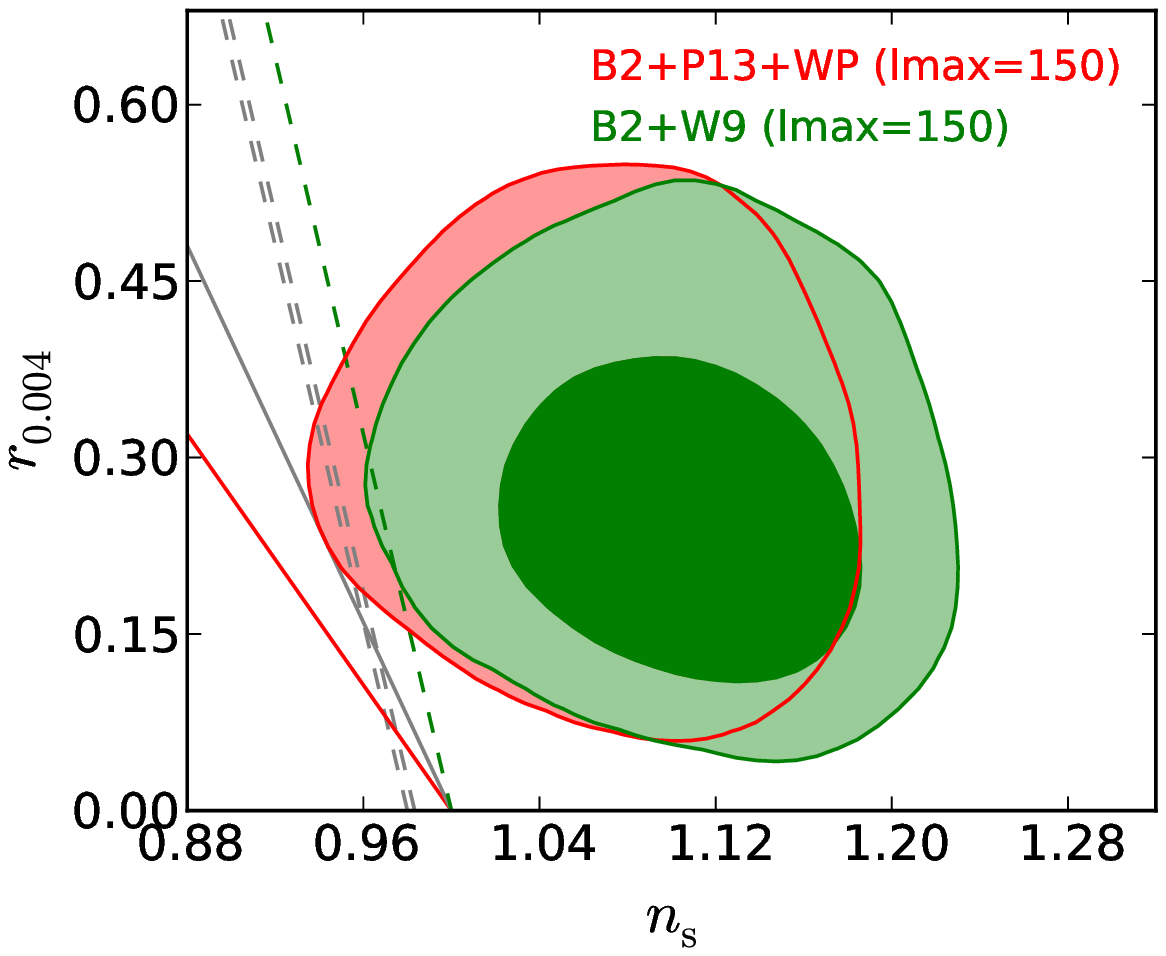}
\includegraphics[width=\figurewidths]{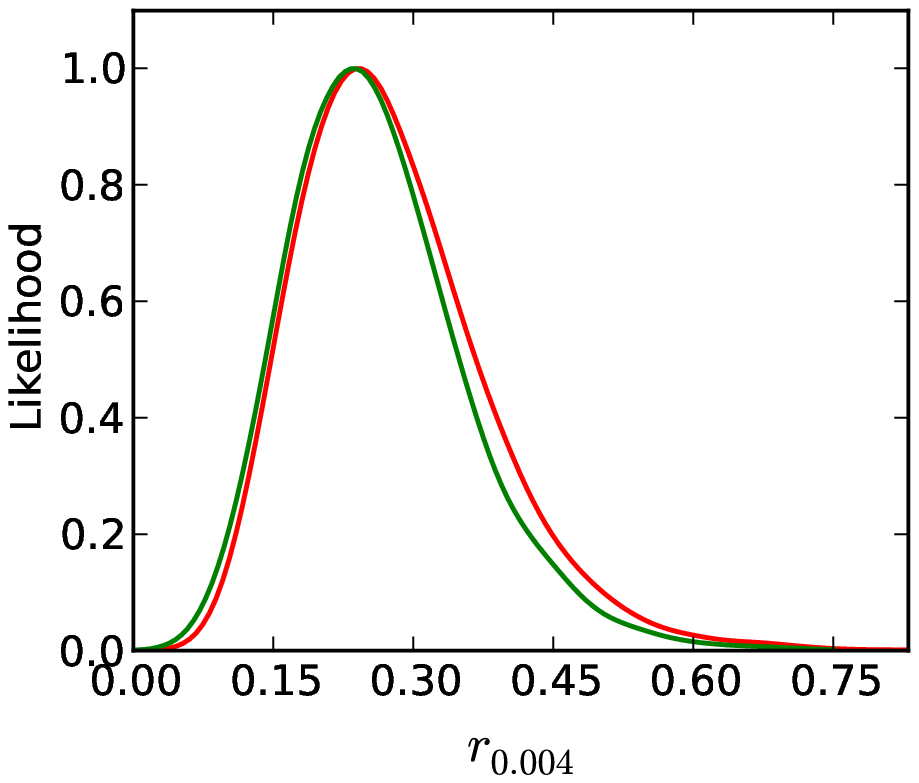}
\includegraphics[width=\figurewidths]{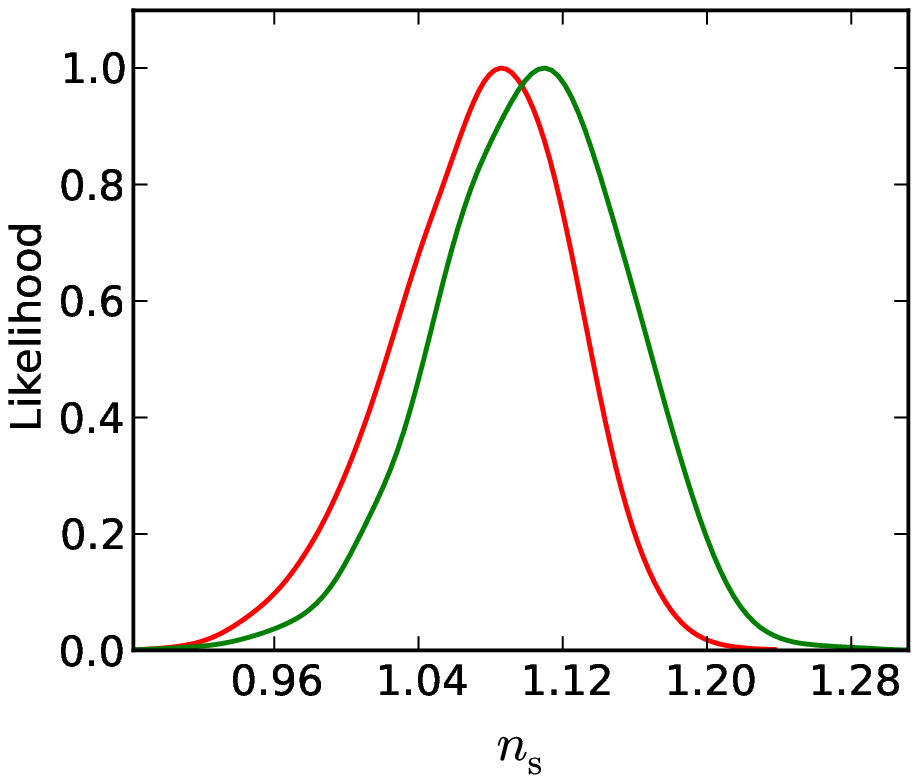}
\caption{The constraints on $r$ and $n_s$ from low-$\ell$ CMB data in the $\Lambda$CDM+r model with $n_t=-r/8$. The red solid line corresponds to inflation with $V(\phi)\sim \phi$. The region between the two gray dashed lines corresponds to e-folding number within $N\in [50,\ 60]$ for inflation models with potential $V(\phi)\sim \phi^n$, and the gray solid line corresponds to $V(\phi)\sim \phi^2$. The green dashed line shows the prediction of the power-law inflation with potential $V(\phi)=V_0 \exp \(-\sqrt{2/ p}{\phi/ M_p}\)$. }
\label{fig:cssi}
\end{figure*}

From now on, let's switch to constrain the canonical single-field slow-roll inflation model in which there is a consistency relation between the tensor-to-scalar ratio $r$ and the tilt of relic gravitational waves spectrum $n_t$, namely $n_t=-r/8$ \cite{Liddle:1992wi}. Therefore here are seven free running parameters: $\{\Omega_bh^2, \Omega_ch^2, \theta, \tau, A_s, n_s, r\}$. Similar to the former case, we also run the CosmoMC \cite{cosmomc} to figure out these cosmological parameters. See Table \ref{tab:cssi} and Fig.~\ref{fig:cssi}. 
\begin{table}[htbp]
\centering
\renewcommand{\arraystretch}{1.5}
\scriptsize 
{
 
\

\begin{tabular}{c|cc|cc}
\hline\hline
$n_t=-r/8$ & \multicolumn{2}{|c|}{B2+W9 $(\ell_{\rm max}=150)$} & \multicolumn{2}{c}{B2+P13+WP $(\ell_{\rm max}=150)$} \\
\hline
parameters& best fit  &$68\%$ limits  &best fit & $68\%$ limits \\
\hline
$\Omega_b h^2$ & 0.0192 & $0.0270_{-0.0104}^{+0.0074}$ & 0.0209 & $0.0263_{-0.0149}^{+0.0077}$ \\
$\Omega_c h^2$ & 0.140 & $0.166_{-0.046}^{+0.029}$ & 0.141 & $0.141_{-0.038}^{+0.018}$\\ 
$100\theta_{\rm MC}$ & 1.116 & $1.104_{-0.048}^{+0.046}$ & 1.142 & $1.098_{-0.027}^{+0.044}$\\
$\tau$ & 0.095 & $0.099_{-0.018}^{+0.015}$ &  0.105 & $0.099_{-0.018}^{+0.015}$ \\
$\ln(10^{10}A_s)$ & 3.106 & $3.051_{-0.054}^{+0.061}$ & 3.021 & $3.024_{-0.057}^{+0.065}$ \\
$n_s$ & 1.098 & $1.104_{-0.051}^{+0.052}$ & 1.120 & $1.074_{-0.042}^{+0.056}$ \\
$r$& 0.19  & $0.26_{-0.11}^{+0.07}$  & 0.23 & $0.28_{-0.12}^{+0.07}$  \\
\hline
\end{tabular}
}
\caption{Constraints on the cosmological parameters from low-$\ell$ CMB data in the $\Lambda$CDM+r model with $n_t=-r/8$.  }
\label{tab:cssi}
\end{table}
The combinations of both B2+W9 and B2+P13+WP with $\ell_{\rm max}=150$ give similar results. 
In \cite{Cheng:2014cja} we used all of data in B2 and W9 to constrain the cosmological parameters. Compared to  \cite{Cheng:2014cja}, we see that there is no tension between the results we get in this letter and those in \cite{Cheng:2014cja}. But here a blue tilted scalar power spectrum is preferred at $2.0\sigma$ level from B2+W9 and at $1.8\sigma$ level from B2+P13+WP respectively if only the low-$\ell$ CMB data $(\ell_{\rm max}=150)$  are adopted. Here we need to stress that the low-$\ell$ CMB data can significantly reduce the possible biases from the complicated microphysics at the recombinations, and the results in this letter directly response to the physics in the primordial Universe.

Here we only consider the CMB data from $\ell=2$ to $\ell=150$ which roughly correspond to $\Delta N=\ln (150/2)\simeq 4.3$ e-folding numbers during inflation. During this short period, inflaton field excuses $|\Delta \phi|/M_p=\sqrt{r/8} \Delta N\simeq 0.7$. Similar to \cite{Cheng:2014cja}, we consider several large field inflation models. Because the contours in the left panel of Fig.~\ref{fig:cssi} stay on the right hand side of the red solid line corresponding to $V(\phi)\sim \phi$, it implies that the potential of inflaton field is convex. The region between two gray dashed lines corresponds to the predictions of chaotic inflation \cite{Linde:1983gd} with potential $V(\phi)\sim \phi^n$ for $n>0$, and the green dashed line corresponds to the prediction of the power-law inflation \cite{Lucchin:1984yf} where the potential of inflaton field goes like $V(\phi)=V_0 \exp \(-\sqrt{2/ p}{\phi/ M_p}\)$. Compared to the constraints from low-$\ell$ CMB data, both the chaotic inflation and the power-law inflation models are marginally disfavored at around $2\sigma$ confidence level. But the inflation model with inverse power-law potential $V(\phi)\sim 1/\phi^n$ for $n>0$ \cite{Cheng:2014cja,Barrow:1993zq} predicts $n_s=1-{n-2\over 8n}r$ which can nicely fit the data if $n<2$.

In addition, space-time is in general non-commutative in string theory \cite{Yoneya:1987,Li:1996rp,Yoneya:2000bt}, namely 
\m
\Delta t \Delta x\gtrsim l_s^2,
\n 
where $l_s=1/M_s$ is the string length scale. 
In \cite{Huang:2003zp,Huang:2003hw,Huang:2003fw}, the effect of space-time non-commutativity makes an extra contribution to the spectral index in the canonical single-field slow-roll inflation model, 
\m 
n_s=1-6\epsilon+2\eta+16\epsilon \mu,
\n 
where $\epsilon={M_p^2\over 2}\({V'\over V}\)^2$ and $\eta=M_p^2{V''\over V}$ are the slow-roll parameters, and $\mu=H^2 p^2/M_s^4$. Here $H$ is the Hubble parameter during inflation and $p=k/a$ is the physical momentum mode of perturbation with comoving Fourier mode $k$. The tensor-to-scalar ratio in the space-time non-commutative inflation is still given by $r=16\epsilon$, and thus the correction to the spectral index from the effect of space-time non-commutativity is $\Delta n_s=+r \mu$. For example, for $r\simeq 0.2$ and $\mu\simeq 1/2$, $\Delta n_s\simeq 0.1$. Both the chaotic inflation and the power-law inflation can generate large tensor perturbations, and they can also fit the data if the effect of space-time non-commutativity is not negligibly small.

To summarize, we propose to adopt the low-$\ell$ CMB data to probe the physics of the primordial Universe. We find that a red-tilted power spectrum of relic gravitational waves perturbation is compatible with the data quite well, but the scalar power spectrum is preferred to be blue-tilted at around $2\sigma$ confidence level.  The constraints on the cosmological parameters from the combination of B2+W9 are roughly the same as those from B2+P13+WP. It is reasonable because the statistic errors in the low-$\ell$ CMB data is dominated by the cosmic variance.

Before closing this letter, we also want to remind the readers that a simple resolution to the tension between B2 and P13 is to take a running spectral index into account in \cite{Cheng:2014bta} where the running of spectral index and furthermore the running of running are taken into account. In both cases the scalar power spectrum on large scales is blue-tilted as well.

\vspace{5mm}
\noindent {\bf Acknowledgments. }
We acknowledge the use of Planck Legacy Archive, ITP and Lenovo
Shenteng 7000 supercomputer in the Supercomputing Center of CAS
for providing computing resources. This work is supported by the project of Knowledge Innovation Program of Chinese Academy of Science and grants from NSFC (grant NO. 10821504, 11322545 and 11335012).



\end{document}